\documentclass[showpacs,preprintnumbers,superscriptaddress]{revtex4}
\usepackage{graphicx}
\usepackage{dcolumn}
\usepackage{bm}
\usepackage{color}
\usepackage{amsmath}
\usepackage{amssymb}
\usepackage{amsfonts}
\usepackage{esint}

\begin{document}

\title{Topological quantum walks in cavity-based quantum networks}

\author{Ya Meng}
\affiliation{State Key Laboratory of Quantum Optics and Quantum Optics Devices, Institute of Laser Spectroscopy, Shanxi University, Taiyuan, Shanxi 030006, China}
\affiliation{Collaborative Innovation Center of Extreme Optics, Shanxi University, Taiyuan, Shanxi 030006, China}

\author{Feng Mei}
\email{meifeng@sxu.edu.cn}
\affiliation{State Key Laboratory of Quantum Optics and Quantum Optics Devices, Institute of Laser Spectroscopy, Shanxi University, Taiyuan, Shanxi 030006, China}
\affiliation{Collaborative Innovation Center of Extreme Optics, Shanxi University, Taiyuan, Shanxi 030006, China}

\author{Gang Chen}
\email{chengang971@163.com}
\affiliation{State Key Laboratory of Quantum Optics and Quantum Optics Devices, Institute of Laser Spectroscopy, Shanxi University, Taiyuan, Shanxi 030006, China}
\affiliation{Collaborative Innovation Center of Extreme Optics, Shanxi University, Taiyuan, Shanxi 030006, China}

\author{Suotang Jia}
\affiliation{State Key Laboratory of Quantum Optics and Quantum Optics Devices, Institute of Laser Spectroscopy, Shanxi University, Taiyuan, Shanxi 030006, China}
\affiliation{Collaborative Innovation Center of Extreme Optics, Shanxi University, Taiyuan, Shanxi 030006, China}

\begin{abstract}
  We present a protocol to implement discrete-time quantum walks and simulate topological insulator phases in cavity-based quantum networks, where the single photon is the quantum walker and the cavity input-output process is employed to realize the state-dependent translation operation. Different topological phases can be simulated through tuning the single-photon polarization rotation angles. We show that both the topological boundary states and topological phase transitions can be directly observed via measuring the final photonic density distribution. Moreover, we also demonstrate that these topological signatures are quite robust to practical imperfections. Our work opens a new prospect using cavity-based quantum networks as quantum simulators to study discrete-time quantum walks and mimic condensed matter physics.
\end{abstract}

\maketitle

\section{Introduction}

Cavity input-output process is one of the basic building blocks in cavity quantum electrodynamics (QED) \cite{Book1,Book2}. It has been widely used in studying quantum optics and cavity-based quantum information processing \cite{Duan1,Xiao1,Lin1,Xue1,Duan2,Fei1,Fei2,Fei3,Wang1,Wang2,Zhang1}. One of seminal protocols in this regard is the Duan-Kimble model \cite{Duan1}, which has been extensively studied in the past years. In this model, a flying single photon has been input into an optical cavity with a single atom trapped inside. When the coupling between the single atom and photon is in the strong coupling regime, this cavity input-output process can function as a atom-photon controlled phase flip gate. Recent experiments have successfully demonstrated the Duan-Kimble model and also the controlled phase flip gates \cite{Rempe1,Rempe2}. The setup in this model can also be used for single-photon transistor \cite{Lukin1} and naturally scaled up to a cavity-based quantum network \cite{Rempe6}, where different cavity-based quantum nodes are connected by the flying photons. These progresses greatly promote the development of cavity-based quantum networks for scalable quantum computation \cite{Rempe1,Rempe2,Rempe6,Rempe3,Rempe4,Rempe5}.

On the other hand, investigating discrete-time quantum walk (DTQW) in various quantum systems has recently attracted a lot of research attentions, including in photons \cite{Photon1,Photon2,Photon3,Photon4,Photon5,Photon6}, cold atoms \cite{Atom1,Atom2} and trapped ions systems \cite{Ions1,Ions2}. Quantum walk is a quantum analog of the classical random walk \cite{Ah}. Because of the coherence of the quantum states, the information propagates at a ballistic rate rather than a diffusive one in the classical random walks \cite{An}. DTQWs also can provide a power tool to realize quantum computation \cite{QC1} and quantum state transfer \cite{TR}. In addition to quantum information science, DTQWs also can function as a versatile quantum simulator for studying quantum diffusion \cite{SG}, Anderson localization \cite{Ander1} and topological phases \cite{KT1,KT2,JK2,JK4,JK3,CC}. For the quantum simulation of topological phases, many recent research attentions have been paid to investigate the topological boundary states via DTQWs in linear optics and optical lattice systems \cite{PhotonTP1,PhotonTP2,PhotonTP3,PhotonTP4,NC1,NC2,Lattice1}. However, the topological features associated with the bulk states are still less explored, including the topological phase transition.

In this paper, motivated by the recent experiments on cavity input-output process, we propose a protocol using a cavity-based quantum network as a quantum simulator to realize a single-photon DTQWs. In this protocol, the state-dependent translation operation which is the basic ingredient for implementing DTQW can be achieved by the cavity input-output process. Based on this DTQW, we further show that a one-dimensional topological phase characterized by a pair of topological winding numbers can be simulated via many steps of quantum walks in a cavity-based quantum network. The topological phase diagrams versus the rotation angles are also given. We further study the topological features of this DTQW, including the topological boundary states and the topological phase transitions. In particular, we illustrate how to design a cavity-based quantum network with two different topological phases and observe the emerged topological boundary states. All the topological phase transition points between different topological phases can be unambiguously measured from the final output photonic density distribution. Our results are also robust to the imperfections in each step of cavity-assisted quantum walk.

This paper is structured as follows. In Sec. 2, we show how to realize a state-dependent translation operation with the cavity input-output process. In Sec. 3, we present a protocol to realize the topological DTQW in a cavity-based quantum network. In Sec. 4, we illustrate how to create and observe the topological boundary states in such network. In Sec. 5, we demonstrate the topological phase transition can also be directly observed in this quantum simulator. In Sec. 6, we give a conclusion to summarize our work.

\section{State-dependent translation via cavity input-output process}

\begin{figure}[tbp]
\centering
\includegraphics[width=10cm]{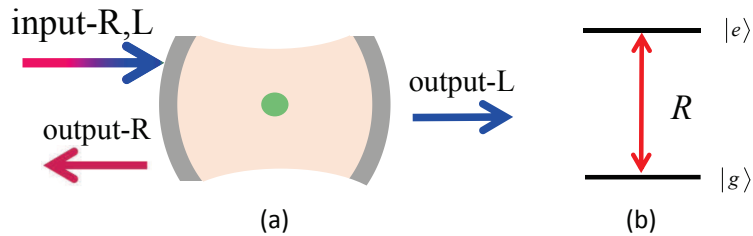}
\caption{(a) Schematic setup for the cavity input-output process. The $L$ component of input photons can go through the cavity but the $R$ component will be reflected. (b) Level structure of the two-level single atom and its coupling with the cavity mode $a_R$.}
\end{figure}

The basic building block in our protocol is the cavity input-output process, which consists of a two-level atom trapped in a two-side optical cavity. The cavity has two resonant modes $a_R$ and $a_L$, with right-circular ($R$) and left-circular ($L$) polarizations, respectively. The input single-photon pulse contains two polarization components $|R\rangle$ and $|L\rangle$. The atomic transition $|g\rangle\leftrightarrow|e\rangle$ is resonantly coupled to the cavity mode $a_R$ and is resonantly driven by the $R$ polarization component of the input single-photon pulse. The $L$ polarization component of the input pulse will see an empty cavity as the atom is decoupled to the cavity mode $a_L$. When the atom is prepared in the state $|g\rangle$, as we will demonstrate below, the $L$ component of the input single-photon pulse will go through the cavity and the $R$ component will be reflected.

In the interaction picture, the interaction of the atom and the cavity mode is given by the Hamiltonian
\begin{equation}
H=g(|e\rangle\langle g| a_R+|g\rangle\langle e| a^\dagger_R),
\end{equation}
where $g$ is the atom-cavity coupling rate. The cavity modes $a_\eta(\eta=R,L)$ are driven by the corresponding input fields $a^{in}_{\eta,l}$ from the left side of the cavity. The Heisenberg-Langevin equations for the cavity modes $a_\eta$ and the atomic operator have the form
\begin{align}
\dot{a}_R&=-ig\sigma^- -\kappa a_R-\sqrt{\kappa}a^{in}_{R,l},\nonumber \\
\dot{a}_L&=-\kappa a_L-\sqrt{\kappa}a^{in}_{L,l},\nonumber \\
\dot{\sigma}^-&=ig\sigma^z a_R,
\end{align}
where $\sigma^-=|g\rangle\langle e|$, $\sigma^z=|e\rangle\langle e|-|g\rangle\langle g|$ and $\kappa$ is the cavity decay rate. The cavity input-output relation connects the output fields $a^{out}_{\eta,j}(j=l,r)$ with the input fields as
\begin{equation}
a^{out}_{\eta,j}=a^{in}_{\eta,j}+\sqrt{\kappa}a_\eta.
\end{equation}
When the cavity decay rate $\kappa$ is sufficiently large, the probability of the atom in the excited state $|e\rangle$ is negligible, the above equations can be analytically solved \cite{Duan1,Duan2}. Under this approximation, we can get
\begin{align}
a^{out}_{\eta,l}&=R_\eta a^{in}_{\eta,l},\nonumber \\
a^{out}_{\eta,r}&=T_\eta a^{in}_{\eta,l},\label{Cavity1}
\end{align}
where the reflection coefficient $R_\eta$ and transmission coefficient $T_\eta$ are given by
\begin{align}
R_R&=|R_R|e^{i\theta_R},T_R=0,\notag\\
T_L&=|T_L|e^{i\theta_T},R_L=0.\label{Cavity2}
\end{align}
In the ideal case, we consider the reflection coefficient $R_R=1$ and the transmission coefficient $T_L=-1$. In this way, the $L$ component of the input single-photon pulse will go through the cavity and acquire a $\pi$ phase shift, while the $R$ component will be reflected. Very recently, this cavity input-output process has been experimentally demonstrated in a single-side cavity with a trapped single atom \cite{Rempe1,Rempe2}. Based on such process, we can realize a polarization-dependent photonic translation.

Different from the discrete time quantum walk using photons in linear optics \cite{PhotonTP1,PhotonTP2,PhotonTP3,PhotonTP4,NC1,NC2}, where the photonic state-dependent translation is generated by classical polariza-tion-dependent optical elements, here the moving of photons is controlled by a quantum atom-photon coupling and its moving direction dependents on both the atomic and photonic internal states, which is important for studying and understanding the coherent feature of quantum walk.
Moreover, the atom-photon interaction can be flexibly tuned in current quantum optics laboratory, which offers more possibilities for designing and studying novel quantum walks. For example, our protocol can be directly generalized to realize a quantum walk with four internal quantum states in the coins by both taking into account the two internal states in the atoms and photons.

\begin{figure}[tbp]
\centering
\includegraphics[width=8cm]{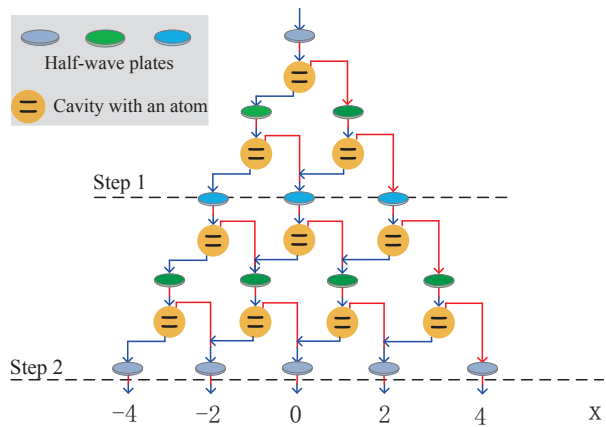}
\caption{Schematic setup for the implementation of DTQW in a cavity-based quantum network. The walker is a single photon with the red (blue) color denoting the  right (left) polarization $R$ ($L$) component of the single photon. The polarization-dependent translation is implemented by the cavity input-output process.}
\end{figure}

\section{Cavity-based topological quantum walks}

In this section, we will show how to realize topological DTQWs. The key operation in a DTQW is the state-dependent translation operation $T$ \cite{KT1,KT2}. In our protocol, this translation can be constructed by the cavity input-output process, which is shown in Fig. 2. Based on Eqs. (\ref{Cavity1}-\ref{Cavity2}), we achieve the polarization-dependent translation operation
\begin{equation}
T=\sum_x|x+1\rangle\langle x|\otimes|R\rangle\langle R|
-|x-1\rangle\langle x|\otimes|L\rangle\langle L|,
\end{equation}
which shows that the $L$ component of the input single-photon pulse will move to the left cavity and acquire a $\pi$ phase shift, while the $R$ component will move to the right cavity. As illustrated in Fig. 2, combined with single-photon polarization rotation operations $R_y(\theta_i)=e^{-i\sigma_y\cdot\frac{\theta_i}{2}}(i=1,2)$, the one-step DTQW operator is written as
\begin{equation}
U(\theta_1,\theta_2)=R_y(\frac{\theta_1}{2})T R_y(\theta_2)T R_y(\frac{\theta_1}{2}),\label{U1}\\
\end{equation}
which is equivalent to the evolution operator generated by a time independent
effective Hamiltonian $H_{\text{eff}}$ over a step time $\delta t$ \cite{KT1,KT2}, i.e. $U=e^{-iH_{\text{eff}}\delta t}$. After $N$ steps of DTQW, the evolution operator becomes $(U)^N=e^{-iH_{\text{eff}} N \delta t}$. In this case, the resulted DTQW simulates the evolution of an effective Hamiltonian $H_{\text{eff}}$ at the discrete times $N\delta t$. In the following, we take the step time of DTQW as $\delta t=1$. By transforming the above Hamiltonian into momentum space, we can get the effective Hamiltonian as
\begin{equation}
H_{\text{eff}}=\int_{-\pi}^\pi dk [E(k) \textbf{n}(k)\cdot\sigma]
\otimes|k\rangle\langle k|,
\end{equation}
where $\sigma=(\sigma_x,\sigma_y,\sigma_z)$ is the Pauli matrices defined on the photonic polarization basis, $E(k)$ and $\textbf{n}(k)$ are the quasi-energies and the unit vector field, respectively. The explicit forms of the eigenvalues and the spinor eigenstates are derived as
\begin{eqnarray}
\cos E(k)&=&\cos\frac{\theta_2}{2}\cos\frac{\theta_1}{2} \cos2k+\sin\frac{\theta_2}{2}\sin\frac{\theta_1}{2},\notag\\
\textbf{n}(k)&=&(0,
\frac{\cos\frac{\theta_2}{2}\sin\frac{\theta_1}{2}\cos2k
-\sin\frac{\theta_2}{2}\cos\frac{\theta_1}{2}}
{\sin E},
-\frac{\cos\frac{\theta_2}{2}\sin2k}
{\sin E}).
\end{eqnarray}

The topological features of DTQWs are characterized by two topological winding numbers $(\nu_0,\nu_{\pi})$ at the quasienergies $E_q=0$ and $E_q=\pi$ \cite{JK2,JK4}. We have numerically calculated the topological phase diagram in Fig. 3 (a). One can find that various topological phases can be prepared via tuning the rotation angles $\theta_1$ and $\theta_2$. The topological phase transition occurs at the gap closing points at $E_q=0$ as well as $E_q=\pi$. In Fig. 3(b), we have also plotted the
the quasienergies $E_q=\pm |E(k)|$ when $\theta_2$ is fixed. The quasienergy gap at $E_q=0$ or $E_q=\pi$ does close in the phase transition points. The number of boundary states at $E_q=0$ ($\pi$) are equal to the difference of the winding numbers $\nu_0$ ($\nu_\pi$) in the two sides of the phase transition points, which yields the bulk-edge correspondence for topological quantum walks.

\begin{figure}[tbp]
\centering
\includegraphics[width=12cm]{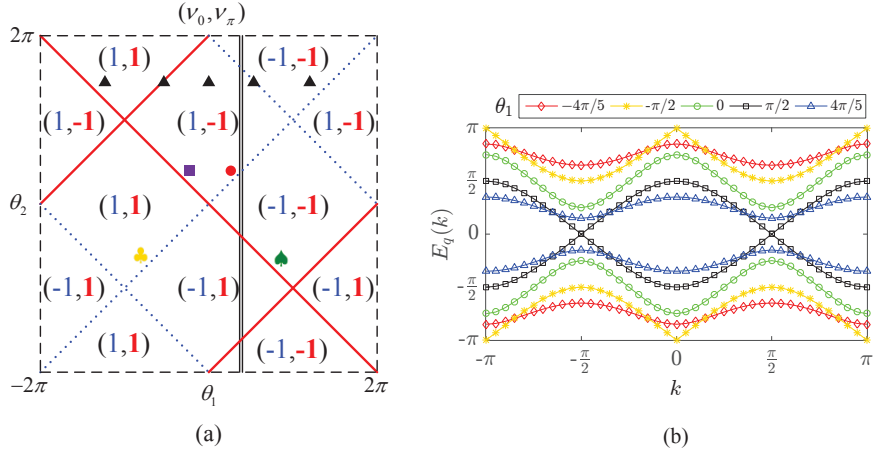}
\caption{ (a) The topological phase diagram of the DTQW $U(\theta_1,\theta_2)$. Different topological phases are characterized by $(\nu_0,\nu_\pi)$. The phase boundaries correspond to points where the quasienergy gap closes at $E_q=0$ (blue dotted lines) and $E_q=\pi$ (red solid lines). The black double lines indicate the rotation angles for observing the topological phase transition. (b) The quasienergy spectrum of the cavity-based DTQW, where the rotation angles $\theta_{1,2}$ are chosen as the values in the five black triangles in (a), with $\theta_2=3\pi/2$.}
\end{figure}

\section{Observation of topological boundary states}

According to the bulk-edge correspondence, boundary states will emerge at the boundaries between different topological phases \cite{TP1,TP2}. Topological boundary states are one of basic signals showing the existence of topological phase. In this section, we will study three cases and show how to observe the topological boundary states generated at the boundaries. The boundary can be created by making the rotation angles $\theta_1,\theta_2$ spatially inhomogeneous in the cavity-based DTQW, such as $(\theta_1^l,\theta_2^l)$ in the left region $x<0$ and $(\theta_1^r,\theta_2^r)$ in the right region $x\geq0$.

In the first case, we consider two DTQW spatial regions with different rotation angles $\theta_1,\theta_2$, i.e. $(\theta_1^l,\theta_2^l)=(-\pi/4,3\pi/8)$ and $(\theta_1^r,\theta_2^r)=(3\pi/4,-5\pi/8)$. As demonstrated in the last section, the topological invariants of the two regions are $(\nu_0,\nu_\pi)=(1,-1)$ and $(\nu_0,\nu_\pi)=(-1,-1)$, respectively. As the topological invariant $\nu_0$ for the two regions are different, we expect to observe the topological boundary states with quasienergy $E_q=0$ near the boundary $x=0$. To experimentally observing such topological boundary states, a single photon pulse with polarization $1/\sqrt{2}(|R\rangle+|L\rangle)$ has been input into the cavity with position $x=0$. Suppose this initial state is denoted as $|\psi(0)\rangle$. After that, as shown in Fig. 2, we implement 15 steps of cavity-based DTQW governed by $U$ and measure the final photon density distribution in the cavity outputs. In Fig. 4 (a), we have numerically calculated the final photon density distribution distribution $P(x,N)$
\begin{equation}
P(x,N)=|\langle x,R|\psi(N)\rangle|^2+|\langle x,L|\psi(N)\rangle|^2,
\end{equation}
where the final state $|\psi(N)\rangle=U^N|\psi(0)\rangle$. If there exists a topological boundary mode around the boundary $x=0$, the input photon at $x=0$ will resonate with this boundary mode and the final photon density will have a peak at $x=0$. Our numerical result in Fig. 4 (a) shows that the photon density distribution after 15 steps of DTQW is nonvanishing around the boundary $x=0$, which indicates that the system has a topological boundary state at $x=0$.

In the second case, we consider two DTQW spatial regions  $(\theta_1^l,\theta_2^l)=(-3\pi/4,-5\pi/8)$ and $(\theta_1^r,\theta_2^r)=(\pi/4,3\pi/8)$. The topological invariants of the two regions are $(\nu_0,\nu_\pi)=(1,1)$ and $(\nu_0,\nu_\pi)=(1,-1)$, respectively. As the topological invariant $\nu_\pi$ for the two regions are different, the topological boundary states with quasienergy $E_q=\pi$  are also expected in this case, which are confirmed by the numerical results shown in Fig. 4(c).

In the third case, we consider creating a boundary between same topological phases, where the topological boundary state will not appear. For this purpose, the polarization angle $\theta_1,\theta_2$ in two DTQW spatial regions are tuned to $(\theta_1^l,\theta_2^l)=(-\pi/4,3\pi/8)$ and $(\theta_1^r,\theta_2^r)=(\pi/4,3\pi/8)$. In this case, both the topological invariants in the two regions are $(\nu_0,\nu_\pi)=(1,-1)$, topological boundary states will not appear in the boundary $x=0$. To demonstrate this point, we prepare the system into the same initial state $|\psi(0)\rangle$ as shown in the first two cases. In Fig 4 (e), the final photon density distribution after 15 steps of DTQW is numerically calculated. The distribution of the DTQW up to 15 steps shows ballistic behavior and no localization is observed around $x=0$. It means that the system has no resonant topological boundary mode in the boundary $x=0$.
\begin{figure}[tbp]
\centering
\includegraphics[width=10cm]{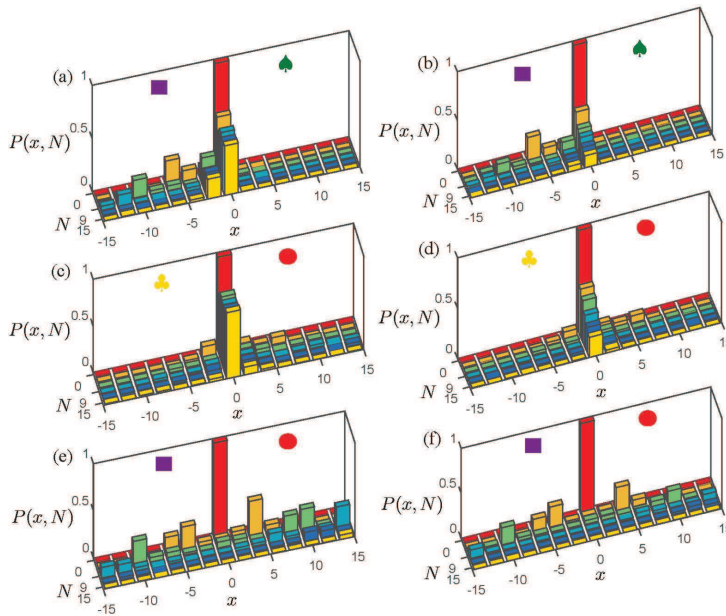}
\caption{The photon density distribution $P(x,N)$ of the inhomogeneous DTQW governed by $U$ in the ideal (a, c, e) and realistic (b, d, f) cases when the number of steps is $N=0$, $3$, $6$, $9$, $12$, $15$. Suppose the single photon walker initially begins at the position $x=0$. The polarization angles are (a-b) $(\theta_1^l,\theta_2^l)=(-\pi/4,3\pi/8)$ and $(\theta_1^r,\theta_2^r)=(3\pi/4,-5\pi/8)$, (c-d) $(\theta_1^l,\theta_2^l)=(-3\pi/4,-5\pi/8)$ and $(\theta_1^r,\theta_2^r)=(\pi/4,3\pi/8)$, (e-f) $(\theta_1^l,\theta_2^l)=(-\pi/4,3\pi/8)$ and $(\theta_1^r,\theta_2^r)=(\pi/4,3\pi/8)$. In the realistic case, the practical parameters in the state-dependent translation operation $T$ are chosen as $R_R=0.98\cdot e^{i\cdot0.05\pi}$ and $T_L=0.98\cdot e^{i\cdot0.95\pi}$, and all the rotation angles have been introduced into a fluctuation $\Delta\in(-\pi/20,\pi/20)$.}
\end{figure}

We also numerically calculate the influence of various imperfections in the cavity input-output process, including the fluctuations of the parameters in $T$ and $R_y$ in each step of the cavity-assisted quantum walk. The numerical results of the photon density distribution are shown in Fig. 4 (b,d,f). For the case supporting topological boundary states, the localization around the boundary of different topological phases decreases because of the loss of the cavity input-output process, but it remains maximal around the boundary. Then we still can unambiguously verify the existence of the topological boundary state even with various imperfections. For the case without topological boundary states, one still can find that there is no photons maximally localized around $x=0$. So, due to the topological protection, the existence of the topological boundary states at the boundaries between different topological phases are robust against small perturbations.

\section{Observation of topological phase transitions}

In this section, we will further show that the topological phase transition between different topological phase can also be directly observed basing on second-order moment associated with final output photon density \cite{NC2}. The rotation angles for observing topological phase transition are chosen as $\theta_1=\pi/3$ and $\theta_2\in[-2\pi,2\pi]$. Similar to the last section, suppose the initial state of the system is prepared into $|\psi(0)\rangle$. To reveal the relationship between the topological phase transition and the final output photon density, the second-order moment is defined as \cite{NC2}
\begin{equation}
M=\sum_x x^2P(x,N)/N^2,
\end{equation}
where $P(x,N)$ is the final output photon density distribution after $N$ steps of DTQW. By transforming the above equation into momentum space, we further get
\begin{equation}
M=\int_{-\pi}^\pi\frac{dk}{2\pi}\langle\psi(0)|
(U^\dagger)^N
(i\frac{d}{dk})^2 (U)^N|\psi(0)\rangle/N^2.
\label{Second}
\end{equation}
The $N$-step evolution operator can be expanded as $(U)^N=e^{-iH_{eff}N}=\cos[N\cdot E]-i\sin[N\cdot E]\textbf{n}\cdot\sigma$. Thus, we can obtain $M$ in the following form
\begin{align}
M&=\int_{-\pi}^\pi\frac{dk}{2\pi}
[\frac{dE}{dk}]^2\notag\\
&+\frac{2i}{N}\int_{-\pi}^\pi\frac{dk}{2\pi}
\{\frac{dE}{dk}\cos[N\cdot E]
\langle\varphi_0|(U^\dagger)^N\frac{d}{dk}
\textbf{n}\cdot\sigma|\varphi_0\rangle\}\notag\\
&+\frac{i}{N}\int_{-\pi}^\pi\frac{dk}{2\pi}
[\frac{d^2E}{dk^2}
\langle\varphi_0|\textbf{n}\cdot\sigma|\varphi_0\rangle]\notag\\
&+\frac{i}{N^2}\int_{-\pi}^\pi\frac{dk}{2\pi}
\{\sin[N\cdot E]\langle\varphi_0|(U^\dagger)^N\frac{d^2}{dk^2}
\textbf{n}\cdot\sigma|\varphi_0\rangle\},\label{X2}
\end{align}
where $|\varphi_0\rangle$ is the initial polarization state of the single photon.

Under the infinite-steps-limit, that is $N\rightarrow\infty$, we ignore these infinitesimal terms in Eq. (\ref{X2}). The form of $M$ becomes particularly simple
\begin{equation}
M=\frac{1}{2\pi}\int_{-\pi}^\pi[\frac{dE}{dk}]^2dk,
\end{equation}
where the quasienergies $E=\arccos(\cos\frac{\theta_2}{2}
\cos\frac{\theta_1}{2}\cos2k
+\sin\frac{\theta_2}{2}\sin\frac{\theta_1}{2})$. The above integral can be rewritten as $M=\oint f(z)dz$ with the complex variable $z=e^{i\cdot k}$, which can be analytically calculated using the residue theorem. After a long straightforward calculation, we obtain
\begin{eqnarray}
M=
\begin{cases}
2, & -2\pi<\theta_2<-\frac{5\pi}{3},\\
4+4\sin\frac{\theta_2}{2}, & -\frac{5\pi}{3}<\theta_2<-\frac{\pi}{3},\\
2, & -\frac{\pi}{3}<\theta_2<\frac{\pi}{3},\\
4-4\sin\frac{\theta_2}{2}, & \frac{\pi}{3}<\theta_2<\frac{5\pi}{3},\\
2, & \frac{5\pi}{3}<\theta_2<2\pi.
\end{cases}\label{Result}
\end{eqnarray}
It turns out that the second-order moment has a plateau when the topological invariants $\nu_\pi=\nu_0$, in contrast to the sine oscillation when the topological invariants $\nu_\pi\neq\nu_0$. Such obvious difference allows us to observe a slope discontinuity at the topological phase transition points in the experiment.

\begin{figure}[tbp]
\centering
\includegraphics[width=10cm]{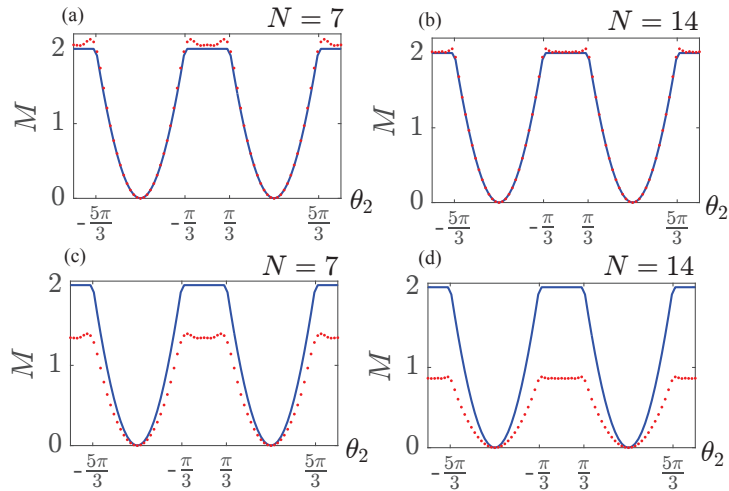}
\caption{The second-order moment varying with different polarization rotation angle $\theta_2$ for the ideal (a-b) and realistic (c-d) cases. The parameters for the realistic case is chosen same as those in Fig. 4. The number of DTQW steps is (a, c) $N=7$, (b, d) $N=14$. The red dots (blue solid lines) denote the numerical (analytical) results.}
\end{figure}

In Fig. 6 (a-b), we have numerically calculated the second-order moment $M$ as functions of the controllable polarization angle $\theta_2$ for different steps of DTQW in the ideal case. It is found that the second-order moment $M$ has a plateaus in the regions $\{-2\pi,-5\pi/3\}$, $\{-\pi/3,\pi/3\}$ and $\{5\pi/3,2\pi\}$. According to the phase diagram, the topological invariants governed by cavity-based DTQW ($U$) in these regions are $(\nu_0,\nu_\pi)=(-1,-1)$. In contrast, $M$ has sine oscillations in the other regions where the topological invariants are $(\nu_0,\nu_\pi)=(-1,1)$ or $(1,-1)$ . Then the phase transition between different topological phases can be clearly observed from the slope discontinuity of second-order moment. We also show that $M$ agrees very well with the theoretical predicated value in Eq. (\ref{Result}) when the number of quantum walk steps $N$ become very large. In Fig. 6 (c-d), we also calculate the influence of the cavity loss in each step of quantum walk on the above results. It turns out that, although the value of the plateaus changes, the plateaus remains in the presence of small imperfections. This feature has not been reported previously \cite{NC2}. It shows that the second-order moment is quite robust imperfections and also has a topological protection.

\section{Conclusion}

In summary, we have proposed a protocol to implement DTQWs in cavity-based quantum networks. Cavity input-output process is employed to realize the state-dependent translation operation,
which recently has been extensively studied in the quantum optics laboratory for implementing large-scale quantum network and scalable quantum computation \cite{Rempe1,Rempe2}. We have shown how to employ cavity-based DTQWs as quantum simulators to mimic and explore the topological phases, including the topological boundary states and topological phase transitions. Our work connects cavity-based quantum computation network with quantum simulation and can motivate more further studies on quantum simulation of condensed-matter physics in this quantum platform.

\section*{Funding}

National Key R$\&$D Program of China (2017YFA0304203); Natural National Science Foundation of China (NSFC) (11604392, 11674200, 11434007); Changjiang Scholars and Innovative Research Team in University
of Ministry of Education of China (PCSIRT) (IRT$\_$17R70); Fund for Shanxi ¡°1331 Project¡± Key Subjects Construction; 111 Project (D18001).

\end{document}